\documentclass[aps,pra,showpacs,onecolumn]{revtex4}
\usepackage{graphicx}

\usepackage{amssymb}
\usepackage{amsmath}
\usepackage{braket}
\usepackage{xcolor}
\usepackage{hyperref}
\usepackage{color}
\usepackage{pbox}
\usepackage{float}
\usepackage{enumerate}

\newcounter{nota}
\newcommand{\nota}[1]{ \stepcounter{nota}
                 \noindent{\bf Nota \Roman{nota}: }  }

\begin{document}

\title{The scaling law of the arrival  time of spin systems that present pretty good transmission}

\author{Pablo Serra$^{(1)}$}
\email{pablo.serra@unc.edu.ar}
\author{Alejandro Ferrón$^{(2)}$}
\email{aferron@exa.unne.edu.ar}
\author{Omar Osenda$^{(1)}$ }
\email{osenda@famaf.unc.edu.ar}

\affiliation{
(1) Instituto de F\'sica Enrique Gaviola (CONICET-UNC) and Facultad de 
Matem\'atica, Astronom\'ia, F\'isica y Computaci\'on, Universidad Nacional de 
C\'ordoba,
Av. Medina Allende s/n, Ciudad Universitaria, CP:X5000HUA C\'ordoba, Argentina\\
(2) Instituto de Modelado e Innovaci\'on Tecnol\'ogica
(CONICET-UNNE) and
Facultad de Ciencias Exactas, Naturales y Agrimensura, Universidad Nacional
del Nordeste, Avenida Libertad 5400, W3404AAS Corrientes, Argentina.
}

\date{\today}

\begin{abstract}
The pretty good transmission scenario implies that the probability of sending one excitation from one extreme of a spin chain to the other can reach values arbitrarily close to the unity just by waiting a time long enough. The conditions that ensure the appearance of this scenario are known for chains with different interactions and lengths.   Sufficient conditions for the presence of pretty good transmission depend on the spectrum of the Hamiltonian of the spin chain. Some works suggest that the time $t_{\varepsilon}$ at which the pretty good transmission takes place scales as $1/(|\varepsilon|)^{f(N)}$, where $\varepsilon$ is the difference between the probability that a single excitation propagates from one extreme of the chain to the other and the unity, while $f(N)$ is an unknown function of the chain length. In this paper, we show that the exponent is not a simple function of the chain length but a power law of the number of linearly independent irrational eigenvalues of the one-excitation block of the Hamiltonian that enter into the expression of the probability of transmission of one excitation. We explicitly provide examples of a chain showing that the exponent changes when the couplings between the spins change while the length remains fixed. For centrosymmetric spin chains the exponent is at most $N/2$. 
\end{abstract}

\maketitle

\section{introduction}\label{sec:introduction}

The transmission of quantum states through spin chains, graphs, or two- and three-dimensional arrays, continues to attract intense scrutiny from theorists and experimentalists alike \cite{Chang2023, Keele2022, Xie2023, Maleki2021, Serra2022, Mograby2021}. Of course, this obeys the double driving exerted by the new experimental setups available and the requirements of a communication channel that must be fast, reliable, and simple to control. 

The homogeneous spin chains with Heisenberg and XX Hamiltonians played a towering role when the field started to call the attention of many researchers \cite{Bose2003, Christandl2004, Christandl2005, Bose-review, Nikolopoulos2015}. There are obvious reasons that justify this early prominence, they both are integrable, so they have complete analytic solutions. Besides, there are well-known numerical methods to treat them when their non-autonomous dynamic behavior needs to be studied. 

Chains with the XX Hamiltonian have a clear advantage over those with the Heisenberg Hamiltonian. The XX chains present perfect transmission without any external fields and for different configurations of the exchange coupling coefficients \cite{Christandl2004, Christandl2005}. They are also easier to control using external time-dependent pulses (see Reference \cite{Coden2021} and references therein). On the other hand, the Heisenberg Hamiltonian, or XXZ Hamiltonians, are  effective Hamiltonians that model most experimental situations \cite{Kandel2021, Martins2017, Kostak2007, quantum-dot-chain, Li2018, Loft2011, Banchi2011prl, Chapman2016, Kandel2019, Baum2021}. 

The controllability of the Heisenberg model \cite{Jurdjevic1972, Burgarth2009, Ramakrishna1995} led to numerous works that dealt with the non-autonomous transmission of states. For homogeneous chains, the resulting performance is somewhat poor. While the controlled dynamics achieve very high values for the transmission fidelities, the designed control pulses are hard to implement, valid only for a specific chain length, the arrival time is too long, and the control pulses change very quickly over time and in a rough way \cite{Wang2016, Burgarth2010, Yang2010, Zhang2016, Farooq2015}.

Given the arguments in the paragraph above, it is no wonder that many studies still focus on transmission protocols based on the autonomous time evolution of the spin chain. The pretty good transmission (PGT) scenario ranks amongst these protocols \cite{Burgarth2006, Godsil2012a, Godsil2012b, Vinet2012}. Roughly speaking, the time evolution of the chain achieves pretty good transmission by waiting for a long enough time. For given conditions, the probability that one excitation transfers from one extreme of the chain to the other is a quasi-periodic function, and it could reach values arbitrarily close to the unity for a given time. 

What distinguishes the PGT setting from other transmission scenarios, as the near-perfect transmission \cite{Zwick3, Banchi2010, Banchi2011} or perfect transmission \cite{Christandl2004},  is that there is a one-to-one relationship between how close to the unity the value of the transmission probability must reach and the time necessary to obtain that closeness. This relationship implies that for any $\varepsilon>0$, there is a time $t_{\varepsilon}$ such that $P(t_{\varepsilon})=1-\varepsilon$, where $P(t)$ is the transmission probability for an excitation transferring from one extreme of the chain to the other. 

The conditions that favor the appearance of PGT  are hard to analyze because of the convoluted relationship between the eigenvalues of the one-excitation Hamiltonian and the exchange coupling coefficients, and sometimes, numerical analysis replaces physical intuition \cite{Kay2019, Kay2010, Banchi2017, vanBommel2010}. Despite the difficulties, the knowledge about the PGT phenomenon has increased in the last few years \cite{Vinet2012,Kay2010,Kirkland2022,Chan2023}. 

On the other hand, while the spectral conditions that guarantee the presence of PGT are well understood, the proper understanding of the scaling of the PGT time with the tolerance $\varepsilon$ and the length of the chain lags behind \cite{Godsil2012a, Vinet2012, Serra2022}. Several attempts to pinpoint the relationship always have found counterexamples. In this paper, we propose that $t_{\varepsilon} \propto \varepsilon^{-\alpha}$, with the exponent $\alpha$ given by the number of linearly independent irrational numbers that enter into the expression of the probability of transmission of
one excitation. The irrational numbers considered are linearly independent over the set of rational numbers, this condition is also calleds rationally independent.  Besides, we show that the XY, homogeneous Heisenberg, staggered Heisenberg \cite{Serra2022}, and decorated SSH chains \cite{Wang2022} show this behavior when PGT is present. All these chains, but the decorated  SSH, have been studied in the PGT context, and we will keep the details about these chains to a minimum while we pay more attention to the decorated SSH chain. 

The SSH chains have increasingly attracted attention because they show localized states with topological properties \cite{Estarellas2017}. The presence of topological, or non-topological, localized eigenstates on the extremes of the chain results in very high transmission fidelities \cite{Estarellas2017, Lemonde2019, Mistadikis2020, Yousefjani2021, Yousefjani2021b}. Moreover, SSH chains are related to a number of experimental implementations, such as transmon chains  \cite{ Wang2022} or QED setups. \cite{Vega2023, Da-Wei2023}. 

We organized the paper as follows. The notation and proposition for the scaling of $t_{\varepsilon}$ with $\varepsilon$ is presented in Section II. The results for XY and Heisenberg chains are presented in Section III, while those concerning the SSH chain belong to Section IV. In Section V, we summarize and discuss our results. 

\section{The scaling of $t_{\varepsilon}$ } \label{sec:the-proposition}

The transmission of a single excitation between sites (or spins) of chains or graphs is the setting where the PGT is studied. Using the computational basis, the quantum state that describes a single excitation localized in the site $i$ of the chain/graph  is denoted by
\begin{equation}
|\mathbf{i}\rangle = |0\rangle_1 \otimes |0\rangle_2 \otimes \ldots |1\rangle_i \otimes |0\rangle_{i+1} \ldots|0\rangle_N,
\end{equation}
where $|0\rangle_j$ and $|1\rangle_j$ are the basis vectors of the Hilbert space associated with the site $j$, $1\leq i \leq N$, and $N$ is the number of sites in the system. 

In the simplest transmission protocol, the initial state of the system is  the quantum state of a single excitation localized at one site
\begin{equation}
|\psi(t=0)\rangle = |\mathbf{i}\rangle.
\end{equation}
If the Hamiltonian of the system is time-independent, then the time-dependent quantum state is given by
\begin{equation}
|\psi(t)\rangle = e^{-itH} |\psi(t=0)\rangle,
\end{equation}
where $H$ is the Hamiltonian and $\hbar=1$. The probability of transmission from site $i$ to site $j$ at time $t$ is given by
\begin{equation}
P_{i,j}^{(N)} = \left| \langle \mathbf{i} | e^{-itH} | \mathbf{j} \rangle  \right|^2. 
\end{equation}

For systems where the total magnetization in the $z$ direction commutes with the Hamiltonian, the probability of transmission reduces to
\begin{equation}
P_{i,j}^{(N)} = \left| \langle \mathbf{i} | e^{-it h_N} | \mathbf{j} \rangle  \right|^2, 
\end{equation}
where $h_N$ is the one excitation block Hamiltonian, given by a $N\times N$ Hermitian matrix \cite{Serra2022}. 

As chains are simple graphs, we use the same notation of  Reference~\cite{Bose2009}. A simple graph, $G(V_n, E_m)$,  is given by two sets, $V_n$, and  $E_m$, which correspond to the set of $n$ vertices and $m$ links, respectively. It is usual to define two matrices $A(G)$ and $J(G)$ such that

\begin{equation}
\label{eag}
\left[ A(G) \right]_{i,j} \,=\,\left\{ \begin{array}{rl}
1  & \mbox{if } (i,j)\,\in \,E_m  \\
0&  \mbox{if not }
\end{array} \right.  \;\mbox{ ;  and }
\left[ J(G) \right]_{i,j} \,=\,\frac{1}{2}\,J_{i,j}\,A(G)_{i,j}
\, .
\end{equation}

\vspace{2ex}

\noindent {\bf The proposition.}

\vspace{1ex}

For a given graph of $N$ spins, with one-excitation Hamiltonian $h_N$, let it be

\begin{equation}
\label{ePc}
P_{i,j}^{(N)}(x_1,\ldots,x_n;t)=\left|\langle i \left|e^{-i\,h_N\,t}\right| 
j\rangle \right|^2\,,
\end{equation}

\noindent the probability that an excitation initially located at site $i$ transfers to site $j$ at time $t$, the quantities $\lbrace x_l \rbrace$ are univocally given by the eigenvalues of $h_N$,  and $|k\rangle$ is the basis of the one-excitation sub-space. 

Assume that there is PGT between the sites $i$ and $j$, then $P_{i,j}^{(N)}$ can be written as

\begin{equation}
\label{ePc2}
P_{i,j}^{(N)}(x_1,\ldots,x_n;t)=
\left(\sum_{i=1}^m\,a_i\,cos(x_i t) \right)^2 \,,
\end{equation}

\noindent where at least one of the $x_i$ is an irrational number, and the following condition holds

\begin{equation}
\sum_{i=1}^m\,|a_i|=1 .
\end{equation}

If there are at most $K_I \leqslant m$ numbers $\lbrace x_i\rbrace$ such that $\lbrace x_1,\ldots, x_{K_I} \rbrace$ are   linearly independent in $\mathbb{Q}$, then

\begin{equation}
\label{esN8}
t_\varepsilon \sim \varepsilon^{-K_I/2} \; .
\end{equation}

\vspace{2ex}

\noindent {\bf Proof.}

\vspace{1ex}

Let us separate the set of numbers $\lbrace x_i\rbrace$ into two subsets, one containing $K_I$ irrationals that are linearly independent on $\mathbb{Q}$, and the other containing $m-K_I$ irrationals linearly dependent on $\mathbb{Q}$ and possibly some rationals. 

Accordingly, with Dirichlet's Theorem \cite{hardy} for a given set of numbers $\xi_1, \ldots, \xi_k \,\in \mathbb{R_+}$\textbackslash $\mathbb{Q}$ the system of inequalities
\begin{equation}
\label{eidat}
|q\,\xi_i-p_i|<\frac{1}{q^{1/k}}\quad;\quad
q,\,p_i \,\in\, \mathbb{N}\,;i=1,\ldots,k
\end{equation}
 has infinity solutions.

Then, we construct approximations for all the $m$ numbers $\lbrace x_i\rbrace$. The fact that there are infinitely many solutions to Eq.~\eqref{eidat} implies that $q$ has not an upper bound. The transmission probability evaluated at time $t=q\pi$ is given by
\begin{eqnarray}
\label{ePcev}
P_{ij}^{(N)}(x_1,\ldots,x_n;t=q\,\pi)=
\left(\sum_{i=1}^m\,a_i\,cos(x_i \,q\,\pi) \right)^2 \simeq
\left(\sum_{i=1}^m\,a_i\,cos[(p_i+1/q^{1/K_I}) \,\pi] \right)^2\, \simeq
\nonumber \\
\left(1-\sum_{i=1}^m\,|a_i| \frac{\pi^2}{q^{2/K_I}}\,\right)^2\sim
1-\frac{cte}{q^{2/K_I}}=1-cte\,\varepsilon \Rightarrow \varepsilon= \frac{1}{q^{2/K_I}}
\Rightarrow \,t_\varepsilon\sim \varepsilon^{-K_I/2} \;\blacksquare
\end{eqnarray}

The implementation of Dirichlet's Theorem implies that once the $x_i$ are known, that is, when an explicit expression for the transmission probability is at our disposal, we look for natural numbers $p_i$ and $q$ such that  $x_i\simeq p_i/q$ and evaluate $P(t_{\varepsilon}=\pi q)$. After this evaluation, we obtain the tolerance as  $\varepsilon=1 - P(t=t_{\varepsilon}=\pi q)$. The approximation for each irrational is better for larger values of $p_i$ and $q$. So, it is possible to construct a succession of approximations for all the irrational numbers in the expression of the transmission probabilities, each term of the success related to a particular value of the number $q$. The biggest drawback to applying the procedure described is the lack of an efficient algorithm that allows finding the numbers $p_i$ and a common denominator $q$ that approximates several irrational numbers. The running-time of the  numerical algorithm that calculates rational approximations for irrational numbers grows exponentially when $K_I \geq 4$. 

In the following Sections, we will present several examples, proceeding to obtain the explicit expressions for the transmission probability and identifying the irrational numbers involved in them. Then, using the algorithm to find successive approximations for the irrational numbers, we plot values of  $P(t_{\varepsilon}=\pi q)$ versus $\varepsilon$. By fitting the data with a scaling law $1/t_{\varepsilon}^{\alpha}$, we will retrieve a numerical value for the exponent. This exponent should be consistent with our proposition. 

In Reference~\cite{Serra2022}, we showed that a family of staggered Heisenberg chains has conclusive pretty good transmission by implementing Dirichlet's  Theorem, as described above, but could not grasp the scaling law proposed in this paper.

\section{Spin chains with Heisenberg and $XX$ Hamiltonians}

The homogeneous XX and Heisenberg Hamiltonians have received attention in the context of PGT. For XX chains, when the one-excitation Hamiltonian is a persymmetric matrix, the spectrum has $\left[ N/2 \right] $ distinct eigenvalues. The eigenvalues ordered, from larger to smaller, satisfy that,

\begin{equation}
\label{ecevxy}
\lambda_{N+1-i}=-\lambda_i \;;\;i=1, \ldots , \left[\frac{N}{2}\right] \;;\;
\lambda_{(N+1)/2}=0 \mbox{ if $N$ odd}\, ,
\end{equation}
where the $\lambda_i>0$, $\lambda_{N+i-1} < 0$ are the roots of a grade $\left[ N/2 \right] $ polynomial. Together with the properties of the eigenvalues of persymmetric matrices, after some direct algebraic steps, we get that 
\begin{equation}
\label{ec:ePc}
\langle 1 \left|e^{-i\,h_N\,t}\right| N\rangle=a_0\,+\,
\sum_{i=1}^{[N/2]}\,a_i\,cos(\lambda_i t)  \,,
\end{equation}
the coefficients $a_i$ are real constants given in terms of the eigenvectors of the Hamiltonian $h$, and $a_0\neq 0$ only when $N$ is odd. 

The product of the eigenvalues is equal to the determinant of $J(G)$, which limits the number of linearly independent irrational eigenvalues to $ \left[ N/2 \right] $ or $ \left[ N/2 \right] -1$ if the determinant is an irrational number or a rational one, respectively. Another way to arrive at these numbers comes from analyzing the characteristic polynomials whose roots are the eigenvalues. 

The analysis above supports our proposition that the exponent of the scaling is the number of LI irrationals that enter into the expression of the probability of transmission. Regrettably, not all the one-excitation Hamiltonians are persymmetric matrices, so now we aim to put forward numerical tests that support our proposition and will allow us to analyze the transmission in more complicated cases. 

We will proceed as in Reference~\cite{Serra2022}. To fix ideas, let us consider a case where the transmission probability includes only one irrational number. For this irrational number $\xi$, produce successive rational approximations of the form $\xi \simeq p_j/q_j$, $j=1,2,3,\ldots$. Then, it is clear that for time $t_{\varepsilon}^j:=\pi q_j$, the condition $|\cos (\xi t_{\varepsilon}^j)| \simeq 1$ is satisfied. Then, evaluating the transmission probability at the successive times $t_{\varepsilon}^j$, we obtain a succession of values $P(t_{\varepsilon}^j)$, which result in a succession of tolerance values defined by $\varepsilon_j:= 1- P(t_{\varepsilon}^j)$.  Note that this procedure is independent of any assumptions about the number of LI irrational numbers. It only depends on the availability of an explicit analytical solution for the probability. Once we get a succession big enough, we can fit the pairs $(\varepsilon^j,  P(t_{\varepsilon}^j))$ using the function $c/t_{\varepsilon}^{\alpha}$, with $c$ and $\alpha$ constants. The fitting exponent should give a value close to $K_i/2$.

In Reference~\cite{Serra2022}, we applied the methodology described in the paragraph above to show that staggered Heisenberg chains show conclusive PGT. The method is limited only to the ability to produce rational approximations for several irrational numbers. 

\begin{figure}[hbt]
\includegraphics[width=0.5\linewidth]{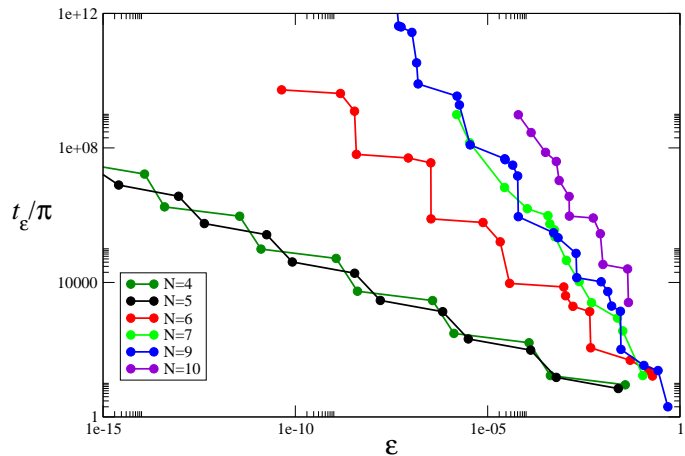}
\caption{The PGT time as a function of $\varepsilon$. The 
data in the figure was calculated for $XX$ chains with lengths $N=4,5,6,7,9$ 
and $10$. Each data point correspond to a pair $(\varepsilon_j, t_{\varepsilon}^j)$, where $\varepsilon_j =1-P(t_{\varepsilon}^j)$. Note that the curves corresponding to $N=7$ and $9$ show the same slope and, correspondingly, the fitting exponent is the same for both cases, see Table~\ref{txy}.
 }\label{fxyscaling} 
\end{figure}

\begin{table}[h!]
\begin{center}
\begin{tabular}{c|c|c}
N&  $K_I$& fitting exponent \\ \hline
4 & 1 & 0.49134 \\
5&1&0.50031\\
6& 2 &0.94557 \\
7&3&1.5067 \\
9&3 &1.5349\\
10& 4 & 1.8185
\end{tabular}
\end{center}
\caption{The Table shows the number of LI irrational numbers, $K_I$, and the fitting exponents obtained for homogeneous $XX$ chains with lengths $N=4,5,6,7,9$, and $10$. The lengths are tabulated in the leftmost column. We fitted the data in Figure~\ref{fxyscaling} using a power law of the form $t_{\varepsilon} \propto 1/(\varepsilon)^{\alpha}$. The exponents obtained from the fitting are tabulated on the rightmost column of the Table.    \label{txy}}
\end{table}

Figure~\ref{fxyscaling} shows pairs $(\varepsilon^j,  P(t_{\varepsilon}^j))$ obtained for chains with lengths $N=4,5,6,7,9$ and $10$. We fitted the data sets for each length and show the fitting exponent in Table~\ref{txy}. The agreement between the numerical value and $K_I/2$ is excellent for all the chains with $N\leq 9$ and, to some extent, only good for $N=10$. The agreement is not so good because of the inherent difficulties in calculating rational approximations with larger and larger denominators. Note that the number of pairs found for $N=10$ is smaller than for the other chain lengths. 

It is worth mentioning that Godsil and collaborators presented the conditions necessary for the appearance of PGT for this type of chain in Reference~\cite{Godsil2012a}. They showed that there is PGT for all the chains with $N\leq 10$, except for the chain with $N=8$. 

The irrational eigenvalues for the chain with $N=7$ are $\sqrt{2}, \sqrt{2\pm \sqrt{2}}$, while for $N=9$ the eigenvalues are $(\sqrt{5}\pm 1)/2$ and $\sqrt{5\pm \sqrt{5}}/2$. Note that in both cases, $N=7$ and 9, there are only three LI irrationals, which explains why the corresponding data sets in Figure 1 have the same slope and that the fitting exponents are indistinguishable.

\vspace{2ex}

\noindent {\bf Staggered Heisenberg spin chains}

\vspace{2ex}

In Reference~\cite{Serra2022}, we dealt with the quantum state transmission problem in chains with Hamiltonian given by
\begin{equation}
H = \sum_i J_i \left( \sigma_i^x \sigma_{i+1}^x + \sigma_i^y \sigma_{i+1}^y +
\sigma_i^z \sigma_{i+1}^z\right),
\end{equation}
where $J_{2l+1}=J_1, l=0, 1,2,\ldots$, and $J_{2l}=J_2, l=1,2,\ldots$.
In the one excitation sub-space, the problem is 
exactly solvable, and  we presented the analytical expressions for the spectrum and eigenvalues 
for centrosymmetric spin chains. Moreover, using some results valid for trigonometric functions, we presented algebraic expressions for the spectrum and eigenvectors of chains with lengths given by
\begin{equation}
N= \alpha 2^k, 
\end{equation}
where $\alpha$ takes one of the values
\begin{equation}
\alpha= \left\lbrace \begin{array}{l}
2 \\
3^p \times 5^q \times 17^r \times 257^s \times 65\, 537^t; \quad  p,q,r,s,t=0,1  \; .
\end{array}\right.
\end{equation}
The algebraic expressions allow the detailed study of the eigenvalues, leading us to conclude that the family of Heisenberg chains that show PGT includes the homogeneous chains and the staggered ones. Besides, for the staggered chains their length must not be restricted to be powers of two. 

\begin{figure}[hbt]
\includegraphics[width=0.5\linewidth]{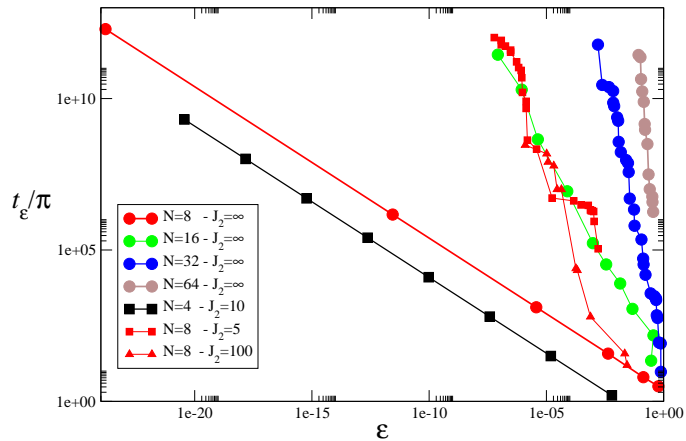}
\caption{The PGT time as a function of $\varepsilon$. The 
data in the figure was calculated for staggered chains with lengths $N=4,8,16,32$ 
and $64$,  which are shown using black, red, green, blue and brown dots, 
respectively. The data points obtained in the strong coupling limit correspond 
to the circular dots, while the data corresponding to finite values of $J_2$ 
are shown using dots with different shapes. For instance, for $N=8$, the data 
for $J=5$ and $100$ are shown using red squares and triangles, respectively. 
}\label{fig:staggered} 
\end{figure}

Here, we summarise our findings for the staggered chains in Figure, \ref{fig:staggered}, and Table, \ref{tf4jpa}. 

\begin{table}[h!]
\begin{center}
\begin{tabular}{c|c|c|c}
N&$J_2$& $K_I$& fitting exponent \\ \hline
8&$\infty$ & 1 & 0.5002 \\
16&$\infty$&3&1.472\\
32&$\infty$&7&3.75 \\
64&$\infty$&15&8.75 \\
4&10&1&0.500\\
8&5&3&1.502\\
8&100&3&1.88
\end{tabular}
\end{center}
\caption{ The Table shows the number of LI irrational numbers enter into the expression of the transmission probability, $K_I$, and the fitting exponents obtained for staggered chains with lengths $N=4,8,16,32$, and $64$. The lengths are tabulated in the leftmost column. We fitted the data in Figure~\ref{fig:staggered} using a power law of the form $t_{\varepsilon} \propto 1/(\varepsilon)^{\alpha}$. The exponents obtained from the fitting are tabulated on the rightmost column of the Table. We included cases  of chains with the same lengths, but different values of the coupling $J_2$ and, in some cases, in the strong coupling limit. This allow us to tune the number of LI irrational numbers, $K_I$. These cases provide a compelling evidence that strengthens our proposition. 
\label{tf4jpa}}
\end{table}

The staggered chains allow for an approximation that simplifies the analysis of the quantum state transfer on large chains. We called this approximation the "strong coupling limit (SCL)'' In this limit, we obtain the probability of transmission using perturbation theory. More importantly, this procedure reduces the complexity of the analytical expression of the transmission probability and, crucially, the number of LI irrational numbers in it. The data sets labeled with $J_2 = \infty$ correspond to the chains studied in that limit. The Figure contains data for chains with lengths $N=4, 8, 16, 32$ and $64$. To further understand the behavior shown by the data in Figure~\ref{fig:staggered}, see Table~\ref{tf4jpa}. In Reference~\cite{Serra2022}, we noted that the behavior of $t_{\varepsilon}$ of a chain  with $N$ spins in the SCL was the same that the one observed for a chain with $2N$ spins with finite coefficients. Compare the data shown for the chains with $N=8, J_2=\infty$ and $N=4, J_2=10$. Moreover, the exponent in those two cases is given by $K_I/2=1$. 

We also fitted the data sets in Figure~\ref{fig:staggered} and included the values of the exponents found in the Table, together with the corresponding value of $K_I/2$. The agreement is excellent, even for the case with $7$ LI irrational numbers. The only large discrepancy appears for the probability of transmission with $K_I=15$. In this last case, the numerical search is too expensive,  and we attribute the difference to the fact that we obtained our numerical data far from the required asymptotic regime.

\section{Decorated SSH chains}

\begin{figure}[hbt]
\includegraphics[width=0.7\linewidth]{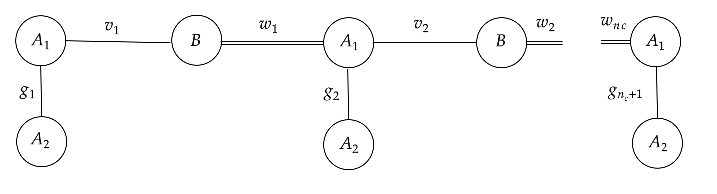}
\caption{The cartoon depicts a decorated SSH chain with three different types of two-level systems, denoted as $A_1$, $A_2$, and $B$. The chain consists of $n_c$ unit cells, each containing three "spins," one of type $A_1$, one of type $A_2$, and one of type $B$. The chain is closed by a last cell that contains only two spins, one of type $A_1$ and one of type $A_2$. The lines connecting the different sites depict the interactions of the Hamiltonian, see Eq.~\eqref{ec:ssh-hamiltonian}. The single lines correspond to the interactions within a unit cell, and the double lines represent the interactions between neighbouring unit cells.  }\label{fig:ssh-decorada}
\end{figure}

The Si-Schrieffer-Heeger (SSH) model \cite{Su1979} is used to study many physical phenomena in condensed matter physics and, more recently, as a model for chains of the superconducting devices known as transmons. In these chains, some qubits control the others, and there are implementations with three different kinds of qubits. The cartoon in Figure~\ref{fig:ssh-decorada} depicts a chain with these characteristics. 

We consider the following Hamiltonian for the decorated SSH chain \cite{Wang2022}
\begin{equation} \label{ec:ssh-hamiltonian}
H= \sum_{i=1}^{n_c} \left( v_i \sigma_{A_{1, i}}^+ \sigma_{B_i}^- + w_i \sigma_{A_{1, i+1}}^ +\sigma_{B_i}^- + \mbox{h.c.} \right) + \sum_{i=1}^{n_c+1} \left( g_i \sigma_{A_{1, i}}^+ \sigma_{A_{2, i}}^-  + \mbox{h.c.} \right) \, ,
\end{equation}
where the strength of the nearest-neighbour interactions $g_i,w_i$, and $v_i$ can be tuned site by site.  It is clear that the chain has a unit cell containing one of each kind of qubit and that the last cell contains only two qubits. If the chain has $n_c$ unit cells, the number of spins in the whole chain is $3\times n_c+2$. The one-excitation Hamiltonian reads as
\begin{equation}\label{ec:ssh-one-excitation}
h = \left(\begin{array}{ccccccccccccc}
     0 &  g_1 &  v_1 &  &  &  &  &  &  &  &  &  & \\
      g_1 & 0 & 0 &  &  &  &  &  &  &  &  &  & \\
      v_1 & 0 & 0 &  w_1 &  &  &  &  &  &  &  &  & \\
     &  &  w_1 & 0 &  g_2 &  v_2 &  &  &  &  &  &  & \\
     &  &  &  g_2 & 0 & 0 &  &  &  &  &  &  & \\
     &  &  &  v_2 & 0 & 0 &  w_2 &  &  &  &  &  & \\
     &  &  &  &  &  w_2 & 0 &  g_3 &  v_3 &  &  &  & \\
     &  &  &  &  &  &  g_3 & 0 & 0 &  &  &  & \\
     &  &  &  &  &  &  v_3 & 0 & 0 &  &  &  & \\
     &  &  &  &  &  &  &  &  & \ddots &  &  & \\
     &  &  &  &  &  &  &  &  &  & 0 &  w_{n - 1} & 0\\
     &  &  &  &  &  &  &  &  &  &  w_{n - 1} & 0 & -g_n\\
     &  &  &  &  &  &  &  &  &  & 0 &  g_n & 0
   \end{array}\right) \, .
\end{equation}

The one-excitation Hamiltonian in Eq.~\eqref{ec:ssh-one-excitation} defines an eigenvalue problem with an exact analytical solution for chains with a small number of unit cells and different sets of interactions. 

Although it is possible to consider the transmission of quantum states from any initial site on the chain to any other final one, we will restrict the present study to the case where the excitation goes from the qubit type $A_1$ of the first unit cell to the qubit type $A_1$ of the last cell. Numbering the qubits of type $A_1$ as 
$1, 4, 7, 3n_c+1$,
of type $A_2$ as
$2, 5, 8, 3n_c+2, $
we will focus on the probabilities of the form
\begin{equation}
P_{1,N-1}^{(N)} , 
\end{equation}
where $N=3n_c+2$. 

\vspace{2ex}

\noindent {\bf SSH chain with  interactions $v=w=1$ and $N=8$}

\vspace{2ex}

The one excitation Hamiltonian is given by
\begin{equation}
h_8 = \left(\begin{array}{cccccccc}
     0 &  g &  1 & 0 & 0 & 0 & 0 & 0\\
      g & 0 & 0 & 0 & 0 & 0 & 0 & 0\\
      1 & 0 & 0 &  1 & 0 & 0 & 0 & 0\\
     0 & 0 &  1 & 0 &  g & 1 & 0 & 0\\
     0 & 0 & 0 &  g & 0 & 0 & 0 & 0\\
     0 & 0 & 0 &  1 & 0 & 0 & 0 & 0\\
     0 & 0 & 0 & 0 & 0 & 0 & 0 &  g\\
     0 & 0 & 0 & 0 & 0 & 0 &  g & 0
   \end{array}\right).
\end{equation}

The eigenvalues are
\begin{equation}
0;0; \pm g; \pm \sqrt{1+g}; \pm \sqrt{3+g} .
\end{equation}

\begin{figure}[hbt]
\includegraphics[width=0.5\linewidth]{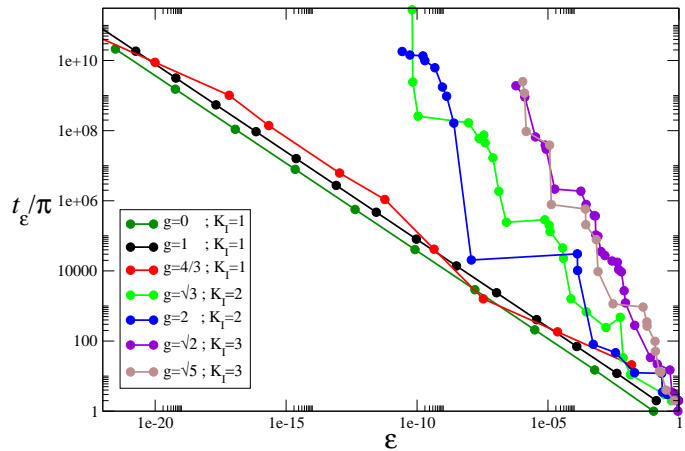}
\caption{The PGT time as a function of $\varepsilon$. The 
data in the figure corresponds to a SSH chain with fixed length $N=8$. The only coupling that is different from the unity is $g$. The slope of the different curves depends on the number of LI irrational number, which we tune precisely to obtain different sets of irrational numbers with $K_I$ elements.  }\label{fig:ssh-n8-g}
\end{figure}

The eigenvectors also have simple analytical expressions, which we use to construct the transmission probability, $P_{1,7}^8$, given by
\begin{equation}\label{ec:16-ssh-analitico}
P_{1,7}^{(8)} = \frac{1}{36} \left(  \cos (\sqrt{3+g^2 }\, t) -3 \cos (\sqrt{1+g^2}\, t) +2 \cos (gt)\right)^2.
\end{equation}
Changing the value of $g$ in Eq. \eqref{ec:16-ssh-analitico}, we can tune the number of linearly independent irrational numbers in the expression above. Table \ref{tab:changin-g-N=8} includes different choices for the value of $g$, the linearly independent irrational numbers on Eq.~\eqref{ec:16-ssh-analitico}, and the exponent found fitting the numerical data in Figure \ref{fig:ssh-n8-g}. 

Figure \ref{fig:ssh-n8-g} shows, for different values of the coupling $g$, the arrival time $t_{\varepsilon}$ versus the tolerance $\varepsilon$. For each value of $g$, the data points correspond to successive approximations calculated for the irrational numbers tabulated in Table \ref{tab:changin-g-N=8}. We fit each data set using a non-linear function $1/t_{\varepsilon}^{\alpha_g}$. The exponents found are those tabulated. With great accuracy, it is clear that $\alpha_g \simeq K_I/2$. 

\begin{table}[h!]
\begin{center}
\begin{tabular}{c|c|c}
g& irrationals & fitting exponent \\ \hline
0 & $\sqrt{3}$ & 0.502 \\
1&$\sqrt{2}$&0.50031\\
4/3&$\sqrt{43}$&0.4969 \\
$\sqrt{3}$&$\sqrt{3},\;\sqrt{6}$&1.0103 \\
2&$\sqrt{5},\;\sqrt{7}$&0.9896\\
$\sqrt{2}$&$\sqrt{2},\;\sqrt{3},\;\sqrt{5}$&1.4443\\
$\sqrt{5}$&$\sqrt{5},\;\sqrt{6},\;\sqrt{8}$&1.3986
\end{tabular}
\end{center}
\caption{The Table shows how many LI irrational numbers,  and the fitting exponents obtained for SSH chains with length $N=8$ and for different values of the coupling $g$, which are tabulated in the leftmost column. We fitted the data in Figure~\ref{fig:ssh-n8-g} using a power law of the form $t_{\varepsilon} \propto 1/(\varepsilon)^{\alpha}$. The exponents obtained from the fitting are tabulated on the rightmost column of the Table. This case clearly shows  that a chain with fixed length, but different values of the coupling $g$, allow us to tune the number of LI irrational numbers, from one up to three. This case provide a compelling evidence that strengthens our proposition. 
\label{tab:changin-g-N=8}}
\end{table}

\vspace{2ex}

\noindent {\bf SSH chains with different lengths with $v=w=1$, $g=0$ or $g=1$}

\vspace{2ex}

Even for moderate lengths, the explicit expression of the transmission probability can be rather cumbersome to write down. We include the results for chains with $N=8, 11, 14$, and $17$ spins. Note that this is equivalent to decorated SSH chains with $n_c=2,3,4$ and $5$. We also include the results for homogeneous XX chains with the same number of spins between the sender and target spins, for comparison purposes since taking $g=0$ in a SSH chain results in an homogeneous $XX$ chain. Note that we will compare the transmission probability of a SSH chain with $N=3n_c+2$ sites with the transmission probability of a $XX$ chain with $2n_c+1$ sites, which we denote as $P_{XX}^{(2n_c+1)}$. 

\vspace{2ex}

\noindent {$\mathbf{N=8}$, $\mathbf{n_c=2}$}

\vspace{2ex}

From Eq.~\ref{ec:16-ssh-analitico}, we that for $g=0$ the transmission probability

\begin{equation}\label{ec:pssh-8-g0}
P_{1,7}^{(8)}(g=0) = P_{XX}^{(5)} = \frac{1}{36} \left(  \cos (\sqrt{3 }\, t) -3 \cos (\sqrt{1}\, t) \right)^2, 
\end{equation}

\noindent while for $g=1$

\begin{equation}\label{eq:pssh-8-g1}
P_{1,7}^{(8)}(g=1)  = \frac{1}{36} \left( \cos (2 t) - 3 \cos (\sqrt{2} t) + 2 \cos (t)
\right)^2 \, .
\end{equation}

\vspace{2ex}

\noindent { $\mathbf{N=11, n_c=3}$ }

\vspace{2ex}

We get that for $g=0 $, the transmission probability is given by
\begin{equation}
\label{epN11g0}
P_{1,10}^{11}(g=0)= P_{XX}^{(7)} = \frac{1}{64} \left[-2 \cos \left(\sqrt{2} t\right)+\left(2+\sqrt{2}\right)
    \cos \left(\sqrt{2-\sqrt{2}} t\right)+\left(2-\sqrt{2}\right) \cos
    \left(\sqrt{2+\sqrt{2}} t\right)-2\right]^2  \, ,
\end{equation}
while for $g=1$ it results
\begin{equation}
P_{1,10}^{(11)}(g=1) = \frac{1}{64} \left[ -2 \cos (t) -2 \cos (\sqrt{3} t) + (2+\sqrt{2}) \cos \left(\sqrt{3-\sqrt{2}} \, t\right) + (2-\sqrt{2}) \cos \left(\sqrt{3+\sqrt{2}} \, t \right)\right] .
\end{equation}

\vspace{2ex}

\noindent {$ \mathbf{N=14,n_c=4}$}

\vspace{2ex}

When we deal with 14 spins we can write, for $g = 0$

\begin{eqnarray}
\label{epN14g0}
P_{1,13}^{14}(g=0) = P_{XX}^{(9)} =
\frac{1}{400} \left[4-\left(5-\sqrt{5}\right) \cos
    \left(\sqrt{\frac{1}{2} \left(3+\sqrt{5}\right)} t\right)
-\left(5+\sqrt{5}\right) \cos \left(\sqrt{\frac{1}{2} 
\left(3-\sqrt{5}\right)} t\right)+ \right. \nonumber \\
\left. \left(3+\sqrt{5}\right) \cos
    \left(\sqrt{\frac{1}{2} \left(5-\sqrt{5}\right)} t\right) 
    +(3-\sqrt{5}) \cos
    \left(\sqrt{\frac{1}{2} \left(5+\sqrt{5}\right)} t\right)
     \right]^2 \,.
\end{eqnarray}

\noindent Note that the two irrational numbers that appear in the arguments of the two cosine functions of the first line of Eq.~\ref{epN14g0} are not LI.

\noindent  For $g=1$ we get 

\begin{eqnarray}
\label{epN14g1}
P_{1,13}^{(14)}(g=1) =
 \frac{1}{400} \left[4 \cos (t)-\left(5+\sqrt{5}\right) \cos
    \left(\sqrt{\frac{1}{2} \left(5-\sqrt{5}\right)} t\right)
-\left(5-\sqrt{5}\right) \cos
   \left(\sqrt{\frac{1}{2} \left(5+\sqrt{5}\right)} t\right)  
\right. \nonumber \\
\left.
+\left(3+\sqrt{5}\right) \cos \left(\sqrt{\frac{1}{2}
    \left(7-\sqrt{5}\right)} t\right)+  
\left(3-\sqrt{5}\right) \cos \left(\sqrt{\frac{1}{2}
    \left(7+\sqrt{5}\right)} t\right)
    \right]^2 \, .
\end{eqnarray}

\vspace{2ex}

\noindent {$\mathbf{N=17,n_c=5}$}

\vspace{2ex}

Finally, for the case of $N = 17$ and $g = 0$, the transmission probability can be expressed as
follows

\begin{eqnarray}
\label{epN17g0}
P_{1,16}^{(17)}(g=0) =  P_{XX}^{(11)} = 
\frac{1}{144} \left[-3 \cos (t)+2 \cos \left(\sqrt{2} t\right)-\cos
    \left(\sqrt{3} t\right)+  \right. \nonumber \\ 
\left. \left(2+\sqrt{3}\right) \cos
    \left(\sqrt{2-\sqrt{3}} t\right)+\left(2-\sqrt{3}\right) \cos
    \left(\sqrt{2+\sqrt{3}} t\right)-2\right]^2 ,
\end{eqnarray}

\noindent and for $g=1$, 

\begin{eqnarray} \label{epN17g1}
P_{1,16}^{(17)}(g=1) = \frac{1}{144} \left[ -2 \cos (t) - \cos (2t) - 3 \cos \left( 
\sqrt{2} \, t \right) + 2 \cos \left( \sqrt{3}\, t \right) \right. + \nonumber \\
 \left. (2+\sqrt{3}) cos \left( \sqrt{3-\sqrt{3}}\, t \right) +
(2-\sqrt{3}) cos \left( \sqrt{3+\sqrt{3}}\, t \right) .
\right]^2
\end{eqnarray}

\begin{figure}[hbt]
\includegraphics[width=0.5\linewidth]{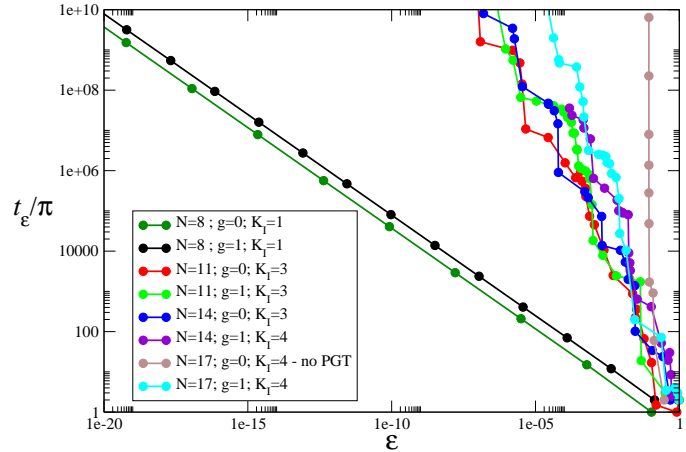}
\caption{The PGT time as a function of $\varepsilon$. The data shown correspond to SSH chains with different lengths $N=8, 11, 14$ and $17$, and two values of the coupling $g=0,1$. To obtain the data we used the transmission probabilities in Eqs.~\ref{ec:pssh-8-g0}-\eqref{epN17g1}. The data in the legend box has the same ordering than the equations. The legend box shows what colour corresponds to each chain length and coupling. The SSH chain with $N=17$ and $g=0$ is equivalent to an $XX$ chain with 11 spins, see Eq.~\eqref{epN17g0}. This chain does not present PGT, see also Reference~\cite{Godsil2012a} }\label{fig:several-n-two-g}
\end{figure}

Figure~\ref{fig:several-n-two-g} and Table~\ref{tab:several-n-two-g} summarize our findings for these decorated SSH and XX chains with $N=8,11,14$ and $17$ spins. 

The Figure shows the data points calculated using rational approximations for the linearly independent irrational numbers from Eq. \eqref{ec:pssh-8-g0} trough Eq.~\eqref{epN17g1}. The  Table shows how many irrational numbers need rational approximations and the exponent obtained fitting the data in the Figure. The agreement between the exponent of the fitting and $K_I/2$ is excellent, except for the case of the longest chain with $g=1$ where the agreement is only good. 

\begin{table}[h!]
\begin{center}
\begin{tabular}{c|c|c|c}
N&g& $K_I$& fitting exponent \\ \hline
8&0 & 1 & 0.502 \\
8&1&1&0.50031\\
11&0&3&1.5039 \\
11&1&3&1.4943 \\
14&0&3 &1.5152\\
14&1&4&1.9804\\
17&0&4&NO PGT\\
17&1&4&2.1082
\end{tabular}
\end{center}
\caption{ The Table shows the number of LI irrational numbers and the fitting exponent for different SSH chains. In the first two columns the Table show the length of the chain and the value of the coupling $g$. By analysing Eqs.~\ref{ec:pssh-8-g0}-\eqref{epN17g1} we get the $K_I$ for each case. Fitting the data in Figure~\ref{fig:several-n-two-g} results in the values tabulated in the rightmost column. The SSH chain with $N=17$ and $g=0$ is equivalent to an $XX$ chain with 11 spins, see Eq.~\eqref{epN17g0}. This chain does not present PGT, see also Reference~\cite{Godsil2012a}.
\label{tab:several-n-two-g}}
\end{table}

\section{Discussion and conclusions}

In actual implementations, the scaling of the arrival time with the length of the chain determines if a protocol is of interest to quantum state transfer or not. In this sense, knowing the conditions a system must fulfil to present PGT is interesting but incomplete. The present work contributes to separating systems with practical interest from systems with arrival times too long to be interesting. The best case is when the expression of the transmission probability contains only one irrational number, but this case only appears for extremely short chains. On a brighter note, our results also show that the number of LI irrational numbers is tuneable by picking adequate couplings in non-homogeneous chains. 

Since perfect quantum state transfer is achievable on chains with couplings tuned to precise values, it seems reasonable that non-homogeneous chains can present PGT with faster arrival times than homogeneous ones.

The assumption that the transmission probability assumes the form of a sum of cosine functions, Eq.~\eqref{ePc2}, could seem restrictive. Nevertheless, all the cases contemplated in this work comply with it.  

The Figures that show the behaviour of $t_{\varepsilon}$ for different systems contain, in some cases, linear curves with equally spaced dots when the vertical scale is logarithmic. These cases correspond to those transmission probabilities whose analytical expression involves only one irrational number. Moreover, sometimes the irrational number to approximate is $\sqrt{n}$, for which the Newton-Raphson recursion relation provides an optimal approximation. For more than one irrational numbers, there is not such an optimal recursion. This drawback explains why there are curves where the density of points becomes scarce or uneven. 

The decorated SSH chains are also known as generalized (or extended) SSH chains. There is complete nomenclature for designing them accordingly with the number of spins in the teeth of the comb-like generalized SSH chains see Figure~\ref{fig:ssh-decorada}.

The one-excitation SSH Hamiltonian, Eq.~\eqref{ec:ssh-one-excitation}, has an analytical solution for many different choices of the coefficients $v_i$, $w_i$, and $g_i$, in particular for the homogeneous chain. If a decorated homogeneous chain with $3k+5$ spins is exactly-solvable, with $k=0,1,2,3, \ldots$, then the chain with $(3k+5) +1 +(3k+5)$ is also exactly-solvable. To generate the larger chain takes gluing two identical chains with $3k+5$ spins using a single B-type spin between the shorter chains \cite{transmon2023}.

The exact transmission probabilities presented for the SSH chains are particular cases of the procedure described in the paragraph above. Because they are a good model and implementable using chains of superconductor qubits \cite{Wang2022} and other QED settings \cite{Vega2023,Da-Wei2023}, there is growing interest in different  SSH chains. More and more evidence shows that these systems, working in the topological or non-topological phase, present PGT, see reference ~\cite{Estarellas2017} and this work. The relevance of PGT for actual experimental implementations seems to be limited, even for chains with topological states whose energies lie in the gap. For excessively long transmission times, even the slightest sources of decoherence could hamper the transmission. We are looking for transmission schemes with short transmission times that do not involve PGT or perturbative analysis. \\

\noindent {\bf Acknowledgements}\\

The authors acknowledge partial financial support from CONICET (PIP 11220210100787CO,
PUE22920170100089CO and PIP11220200100170). AF acknowledges partial financial sup-
port from ANPCyT (PICT2019-0654). OO and PS acknowledges partial financial support from
CONICET and SECYT-UNC.

\appendix

\end{document}